# Real-time Cellular Impedance Monitoring and Imaging of Biological Barriers in a dual flow membrane bioreactor

L. Cacopardo [1,2], J. Costa [1,2], S. Giusti [1,3], L. Buoncompagni [1], S. Meucci [4], A. Corti [1,5], G. Mattei [2] and A. Ahluwalia [1,2]

[1] *Research Centre 'E. Piaggio', University of Pisa, Italy*

[2] *Department of Information Engineering, University of Pisa. Italy*

[3] *IVTech S.r.l, Pisa, Italy*

[4] *Micronit Microtechnologies, Enschede, The Netherlands*

[5] *Department of Translational Research and New Technologies in Medicine and Surgery, University of Pisa, Italy*

**Abstract**

The generation of physiologically relevant in-vitro models of biological barriers can play a key role in understanding human diseases and in the development of more predictive methods for assessing toxicity and drug or nutrient absorption. Here, we present an advanced cell culture system able to mimic the dynamic environment of biological barriers while monitoring cell behaviour through real-time impedance measurements and imaging. It consists of a fluidic device with an apical and a basal flow compartment separated by a semi-permeable membrane. The main features of the device are the integration of a transepithelial electrical impedance (TEEI) meter and transparent windows for optical monitoring within a dual flow system.

Caco-2 cells were cultured in the TEEI bioreactor under both flow and static conditions. Although no differences in the expression of peripheral actin and occludin were visible, the cells in dynamic conditions developed higher impedance values at low frequencies, indicative of a higher paracellular electrical impedance and thus suggesting accelerated barrier and tight junction formation with respect to the static cultures. TEEI measurements at high frequency also enabled monitoring the evolution of transcellular impedance during culture. The cells subject to flow showed a typical RC behaviour, while the controls showed minimal capacitive behaviour, again highlighting the differences between flow and static conditions.

# 1. Introduction

Biological barriers allow the separation between different compartments of the human body or between the body and the external environment. These barriers have a fundamental role in controlling the absorption of exogenous substances such as nutrients and xenobiotics, as well as in the maintenance of homeostasis in different body compartments. Moreover, they act as the first level of defence against microorganisms, toxins, nanomaterials and allergens (Mullin et al., 2005). For these reasons, studying biological barriers is not only crucial for a better understanding of their pathophysiology, but it is also critical in drug testing and toxicology experiments. Epithelial and endothelial cells regulate the passage of substances across these barriers through their junctions and membranes. The cells physically separate two compartments and essentially control the paracellular and transcellular fluxes across the barrier. Indeed, typical in vitro models of biological barriers are based on epithelial cells cultured on the membrane of a Transwell® insert which separates an apical and a basal compartment (Kim et al., 2012; Mullin et al., 2005).

Epithelial or endothelial cell layers (Figure 1A) can be modelled as an electrical parallel between a resistance and a capacitor which respectively represent the paracellular pathway related to tight junctions (TJ) and the transcellular pathway dominated by the capacitance of the hydrophobic phase of the cellular membrane (Benson et al., 2013; Unzel et al., 2012).

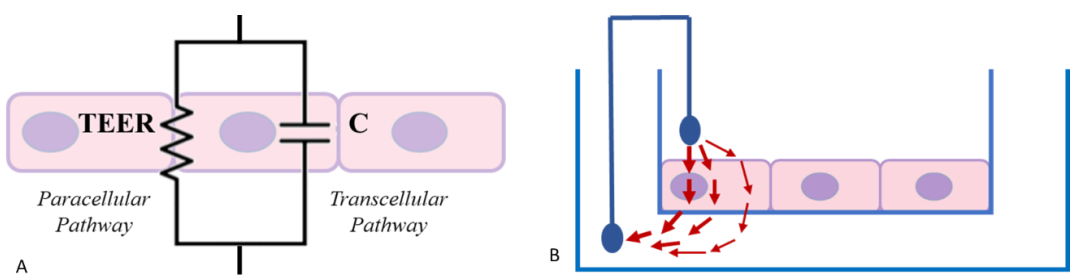

**Figure 1:** A) Electric equivalent of an epithelial or endothelial cell layer; B) Schematization of EVOM (WPI) measurements in transwells with chopstick electrodes (red arrows indicate the electric field distribution).

The integrity of the barrier is usually characterised by measuring the Trans Epithelial Electric Resistance (TEER), which indirectly evaluates the formation of a compact monolayer assessing the formation of TJs. This measurement principle is based on the application of a direct current ($I_{DC}$) and on the recording of the resultant voltage (V), giving information on the ohmic resistance of the cellular layer ($R = V/I_{DC}$). However, TEER measurements are often performed using low frequency alternate currents ($I_{AC}$) rather than direct current (DC) to avoid cell damage and electrode polarization. Although different cell components (e.g. membrane channels) contribute to the ohmic resistance R, the TJ contribution is predominant, thus DC and low frequency measurements mainly evaluate the paracellular electric resistance (i.e. TEER) (Benson et al., 2013; Unzel et al., 2012).

Impedance spectroscopy, i.e. the application of a frequency sweep of current, can provide additional information on the cellular barrier. In particular, at low frequencies (f < 5 kHz), the current flows principally through the paracellular pathway allowing TEER measurements, while at high frequencies it passes capacitively through the membrane (i.e. transcellular pathway), giving information on the cellular adhesion process, migration and micro-motions (Benson et al., 2013). Indeed, ideally when $f \to 0$ the capacitor of the transcellular pathway is fully charged and current cannot flow across it, thus low frequency $I_{AC}$ flows across TEER. Conversely, as the frequency increase the capacitor starts to be progressively conductive (for $f \to \infty$ its impedance tends to zero), i.e. high frequency $I_{AC}$ is able to flow across the cells. This means that the impedance decreases with increasing frequency (Unzel et al., 2012).

The most widely-used TEER measuring system is the Epithelial VoltOhmMeter EVOM (World Precision Instruments – Florida, USA), supplied with chopstick electrodes for measuring the TEER in transwells (Figure 1B). It applies an alternating current (12.5 Hz, 10 µA) to avoid charging of the cell layer and the electrodes. Essentially, the measured resistance includes the resistance of the cellular layer, the porous membrane, the medium and the medium/electrode interface. Each stick of the electrode pair contains a silver/silver-chloride pellet for measuring voltage and a silver electrode for the stimulating current (Benson et al., 2013).

Alternatively, the CellZScope (NanoAnalytics - Münster, Germany), is a computer-controlled multiwell device that automatically derives all the important parameters related to the status of the cellular layer. The electrical stimulus is represented by a 45 mV sinusoidal voltage and a partition of the voltage is used to avoid current read out. The output impedance has three contributions: the TEER, the bulk resistance and the capacitance of the cellular layer. In particular, this last parameter is important for monitoring the growth of cellular extrusions like microvilli (Benson et al., 2013)

On the other hand, in the ECIS (Electrical Cell Substrate Sensing) system (Applied BioPhysics, Inc – NewYork, USA), cells are not grown on porous membranes but directly on integrated gold-film electrodes. The close proximity of the cell monolayer to the thin gold electrodes results in highly sensitive measurements. However, since there is no basolateral fluid compartment due to the adherence of the cells to the electrode, transport and translocation experiments are excluded. The ECIS array is typically composed of 8 wells, each one containing 10 active electrodes (diameter = 250 µm) connected in parallel and a larger counter electrode which is located at the base of the well. Because of the size differences in the electrodes, the measured electrical resistance is mainly determined by the active electrode. Due to the much smaller surface area of the ECIS working electrodes, the measured absolute impedance, formed by the resistance and the capacity on the cellular layer, differs significantly (Benson et al., 2013; Wegener et al., 2000)

Although the most common in-vitro models of epithelial barriers are based on static transwells, several recent studies are focused on the implementation of more physiologically relevant models able to provide cells with an appropriate pattern of physical and chemical stimuli that better reproduce the in-vivo environment. Therefore, barrier-forming cells are often cultured in fluidic systems, or bioreactors, which are able to apply dynamic flow conditions (Giusti et al., 2014; Martin et al., 2004; Sbrana et al., 2016).

Monitoring environmental parameters inside these advanced cell culture systems is critical (Giusti et al., 2017) and some attempts, summarised in Table 1 have been made to determine barrier integrity in bioreactors.

**Table 1:** Overview of TEER and TEEI Measuring Systems and their integration in bioreactors. (*=it depends on the type of transwell or support used)

| Commercial Name or References | TEER | TEEI | Integration with bioreactor | Number of electrodes | Imaging |
|---|---|---|---|---|---|
| *EVOM* | ✓ | | | 4 | * |
| *CellZScope* | | ✓ | | 2 | No |
| *ECIS* | | ✓ | | 2 | * |
| *(Douville et al., 2010)* | | ✓ | ✓ | 2 | No |
| *(Ferrell et al., 2010)* | ✓ | | ✓ | 4 | No |
| *(Vogel et al., 2011)* | ✓ | | ✓ | 2 | No |
| *(Booth and Kim, 2012)* | ✓ | | ✓ | 4 | Yes |
| *(Griep et al., 2013)* | | ✓ | ✓ | 2 | No |
| *(Sbrana et al., 2016)* | ✓ | | ✓ | 4 | No |
| *(Costello et al., 2017)* | ✓ | | ✓ | 4 | No |
| *(Henry et al., 2017)* | | ✓ | ✓ | 4 | Yes |

Douville and co-workers created a microfluidic channel with embedded TEER electrodes in close proximity to the cell monolayer, thus reducing the noise from the electrical resistance of the medium and from electrode motion. Two Ag/AgCl embedded electrodes with a diameter of 500 μm were placed in the side channel and held by elastomeric tension through the sealing process. An alternating current with an amplitude of 0.1 V was applied in the frequency range from 10 Hz to 1 MHz, yielding 64 impedance measurements, recorded through a laboratory potentiostat/galvanostat (Douville et al., 2010). Using microfluidics technology, Ferrel and co-workers designed a bilayer microfluidic system to evaluate kidney epithelial cells under physiologic fluid flow conditions. Ag and Ag/AgCl electrodes were integrated (using tubing and epoxy glue) into the device to monitor cell growth and evaluate the integrity of TJs. In this case, measurements were performed with the EVOM system and the electrodes were connected using a specific adapter (Ferrell et al., 2010). A micro-scaled fluidic system was developed by Vogel and co-workers, allowing the interaction of a vascular cell layer with a flowing stream of red blood cells. An electrode made of an aluminium strip was integrated in the bottom of the channel while a mobile electrode was placed on the top. A bipolar square voltage (0.7 V, 20 Hz) was applied through two electrodes outside the channel and the resulting current was measured and analysed using MATLAB (Mathworks) (Vogel et al., 2011). A multilayer microfluidic device was fabricated by Booth and colleagues using four PDMS substrates, two glass layers, a porous polycarbonate membrane and four AgCl thin-film electrodes, which were then connected to the EVOM (Booth and Kim, 2012). Similarly, a two-layer membrane-based device made of PDMS was proposed by Griep et al. In their system a polycarbonate membrane separated the top channel from the bottom one and platinum wire electrodes with a diameter of 200 μm were placed in contact with the two channels to perform impedance measurements using a spectrum analyser (Griep et al., 2013). Recently, a milli-fluidic bioreactor was integrated with four platinum electrodes deposited in a silicon ultrathin porous support and connected to the EVOM. Continuous TEER measurements were recorded via a NI USB-6008 (National Instrument) acquisition card (Sbrana et al., 2016). Costello et. al developed a microscale bioreactor, which incorporate a scaffold able to mimic the villi topography of the intestine. The bioreactor was realised by 3D printing with a near-transparent material and silver wires were integrated to connect the system with the EVOM (Costello et al., 2017). Finally, in a recent study, a microbioreactor was integrated with four gold electrodes patterned on transparent polycarbonate substrates. Impedance spectra was acquired connecting the electrodes to a PGstat128N from Metrohm Autolab BV (The Netherlands) (Henry et al., 2017).

With the exception of the milli-fluidic device described in (Sbrana et al., 2016), all the bioreactors with integrated TEER/TEEI measurement systems are microfluidic devices and their electrodes are either interfaced with the EVOM or a laboratory impedance analyser, which are not particularly suitable for cellular measurements. Moreover, given their small volumes, micro-fluidic bioreactors have limited space to incorporate multiple electrodes. Indeed, most devices are featured with 2 electrodes only, making it impossible to eliminate solution resistance in the electrochemical cell and thus are more sensitive to any interference coming from junction potentials. Furthermore, only two of these systems (Booth and Kim, 2012; Henry et al., 2017) allow optical monitoring of cells during culture.

To address some of these issues, we developed a new milli-fluidic double-flow bioreactor (called TEEI bioreactor), which integrates a semi-permeable membrane and four electrodes, to perform real-time monitoring of an epithelial cell layer using both impedance spectroscopy and live imaging. The impedance measuring system was purposely designed to be compact and robust and to enable both TEEI and TEER measurements without damaging the cells nor interfering with dynamic experiments. The TEEI bioreactor has several advantages with respect to microfluidic devices, including i) ease of assembly, ii) low surface to volume ratio, iii) absence of any trapped bubbles, iv) low shear stress in presence of fairly high flow rates, v) low pressure gradients and vi) the possibility of seeding a more physiological number of cells (Ahluwalia, 2017; Mattei et al., 2014). This paper describes the design of the TEEI bioreactor and its application to the characterisation of an intestinal barrier model based on Caco-2 cells.

## 2. Materials and Methods

### 2.1 The TEEI Bioreactor

The TEEI bioreactor is an adaptation of the modular, dual flow commercial Live Box 2 bioreactor (LB2, IVTech S.r.l. - Massarosa, Italy). Barrier forming cells are grown on a porous membrane placed between an apical and a basal chamber. Caco-2 cells exposed to 48h of medium flow in a similar system were shown to exhibit improved barrier integrity (Giusti et al., 2014). One of the main features of the LB2 is its optical transparency, which allows performing real-time imaging of the culture environment using a microscope.

Figure 2A illustrates the different components of the TEEI bioreactor. It is composed of an apical and a basal chamber equipped with silicone tubes for culture medium inflow and outflow. A membrane placed in a Teflon (polytetrafluoroethylene, PTFE) holder separates the apical and the basolateral chambers, which are closed by two circular glass slides (20 mm diameter, 150 μm thick), one at the top and one at the bottom of the system, to allow live imaging.

The electrodes (Figure 2B) were fabricated on the inner part of the 150 µm-thick glass slides (i.e. the one in contact with the culture chamber) at Micronit Microtechnologies BV. The conductive structures were created by dispensing a silver enriched epoxy paste (The Gwent Group, UK) with a dispensing CNC (computer numerical control) machine. The circular electrodes were purposely designed to optimize the electric field across the membrane and maximise the field of view for imaging. The printed electrodes are 338±1 µm in thickness and 33±1 µm in height.

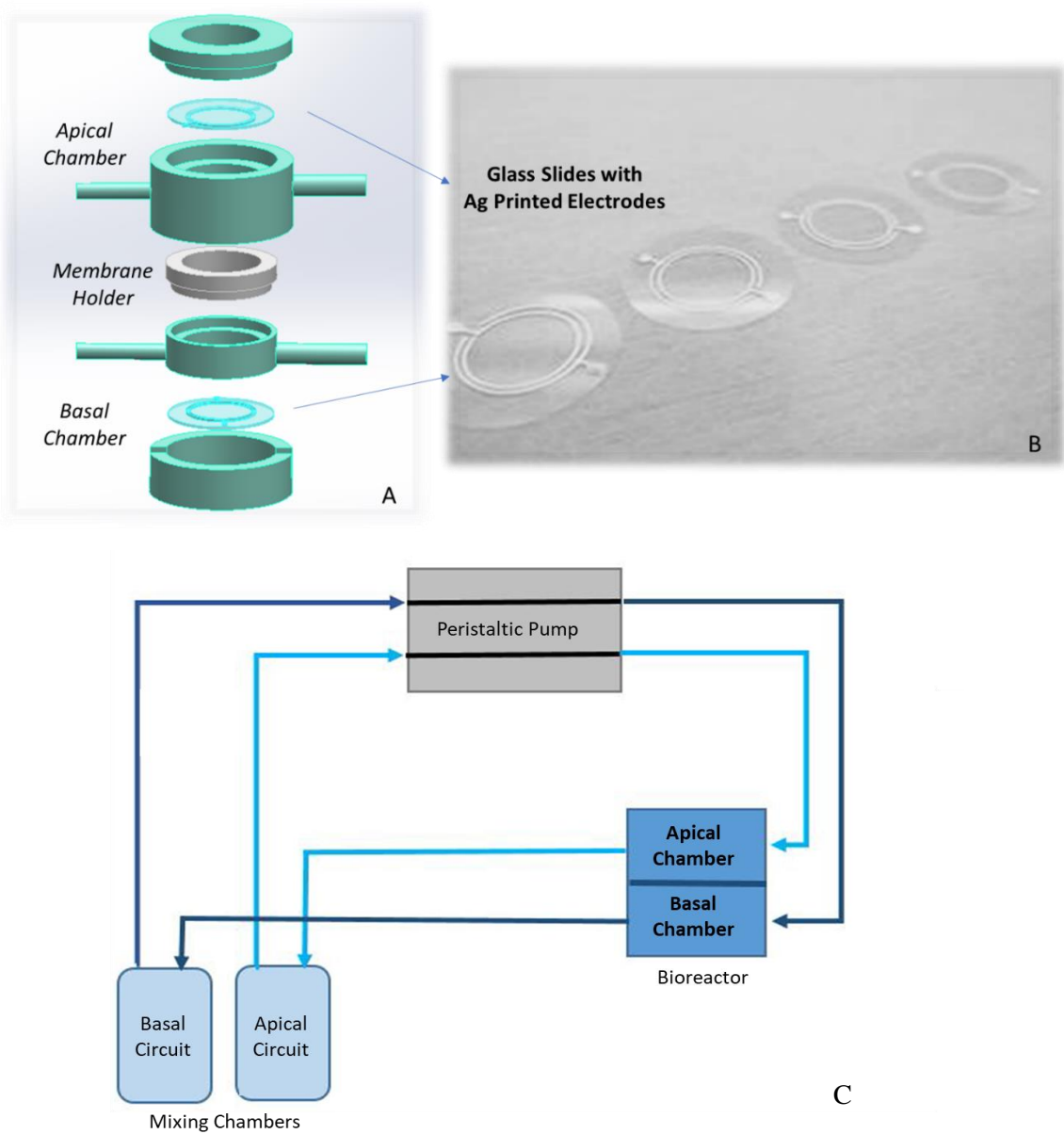

**Figure 2**: A) Schematic view of the TEEI Bioreactor; B) Glass slides with the silver printed electrodes; C) Schematic view of the fluidic circuit: the two separate apical and basal circuits are connected to a peristaltic pump and two mixing chambers acting as fluid as reservoirs.

The dimensions of the TEEI bioreactor culture chamber are similar to those of a well of a 24 multi-well plate (cell seeding area of ~1.8 cm$^2$). Except for the glass slides and the membrane holder, the bioreactor was made of polydimethylsiloxane (PDMS, Sylgard 184 - Dow Corning, Michigan, USA), a biocompatible polymer with good sealing properties. A clamp system (Figure 3C) was purposely designed to guarantee watertight closure by applying a pressure on its vertically-stacked components (Figure 2A) and realise an electrical connection with the printed electrodes in the bioreactor chamber thanks to embedded gold spring contacts. The TEEI bioreactor is connected as a plug and play device to the external impedance measuring system described in section 2.2. The bioreactor clamp and external electrical coupling components were designed with SolidWorks (Massachusetts, USA) and 3D printed in Acrylonitrile-butadiene-styrene (ABS) with a 3D printer (Fortus 250 - Stratasys, Minnesota, USA).

For culturing cells under dynamic conditions, the apical and basal chambers of the TEEI bioreactor are connected to two different mixing chambers via silicone tubing, thus defining two different fluidic circuits, as schematised in Figure 2C.

## 2.2 The cellular Impedance-meter

The cellular impedance measurement circuit is based on the AD5933 (Analog Devices Inc.) and on an analog front-end (AFE) purposely developed to adapt the chip to biological measurements (Margo, 2013) and allow low frequency measurements (Brennan, 2007) resulting in a frequency range from about 40 Hz up to 100kHz. Moreover, a tetra-polar electrode configuration was adopted to avoid artefact impedances at the electrodes interface. A microcontroller (Arduino Micro - Interaction Design Institute, Ivrea, Italy) was used to program the AD5933 chip setting up the initial frequency ($f_i$), the frequency step ($\Delta f$) and the number of steps (N). In addition, measurements can be performed and data downloaded via the serial interface.

The printed circuit board (PCB) was designed using EAGLE (AutoDesk - California, USA). The impedance-meter case contains both the PCB and the power supply, which is composed of a 9 V alternator and a chip (K8042 Velleman – Gavere, Belgium) that ensures a symmetric powering (i.e ± 5 V).

## 2.3 Computational models

A computational fluid dynamic (CFD) model of the TEEI bioreactor was developed in COMSOL Multiphysics 4.3b (Stockholm, Sweden). The viscosity of the culture medium was considered equal to that of water at 37°C (i.e. 0.6913 mPa/s (Kestin et al., 1978).

While, electric field distribution was studied using the Electric Currents module of COMSOL Multiphysics. In particular, a current of 80 µA, equal to stimulation current provided by the front-end was applied to the electrodes. The electric properties of the culture medium (considered an aqueous media at 310.15 K), the cell layer and electrodes are reported in **Error! Reference source not found.Error! Reference source not found.** (Griffiths, 1999; Onnela et al., 2012).

Table 2: Electrical parameters used to model the electric field distribution in the TEEI bioreactor.

|  | **Relative electric permittivity ($\varepsilon_r$)** | **Electric conductivity ($\sigma$)** |
|---|---|---|
| Aqueous media | 80.1 | 1.5 S/m |
| Cellular layer | $1.18 \times 10^6$ | 0.38 S/m |
| Silver electrodes | 0.403 | $1.16 \times 10^8$ S/m |

## 2.4 Impedance measurements

To evaluate the performance of the system (i.e. sensitivity, linearity and stability) and correlate results with those obtained using the EVOM, preliminary tests were performed with sodium chloride solutions at different concentrations (0.001 M, 0.005 M, 0.01 M, 0.1 M and 1 M). For each solution, the impedance magnitude (|Z|) was measured both with our impedance-meter in the bioreactor (at 40 Hz) and with the EVOM in the transwells® (at 12.5 Hz).

## 2.5 Cell Cultures

Although the TEEI bioreactor can be used to study different types of barrier, in this work an in-vitro model of the human intestine was implemented as specific application to test the performance of the device. Caco-2 cells form monolayers with TJs and express a variety of transporter proteins. The standard culture protocol for these cells is based on a 21-day long experiment on semi-permeable supports such as Transwells®, thus being quite expensive both in terms of time and resources. However, 7-day long experiments with Caco-2 cells have been reported to provide similar results to those of traditional 21-day long models in terms of cellular morphology, monolayer integrity, paracellular and transcellular permeability (Cai et al., 2014; Peng et al., 2014). Moreover, it has been demonstrated that culturing Caco-2 cells under flow conditions induces the formation of differentiated epithelial cell monolayers in shorter times than in static conditions (Giusti et al., 2014; Kim et al., 2012).

In this work, Caco-2 cells were cultured using high glucose DMEM, supplemented with 10 % FBS, 1% penicillin/streptomycin, 1% non-essential amino-acids and 1% L-glutamine. Polyethylene terephthalate (PET) membranes (it4ip - Louvain-la-Neuve, Belgium) were used as semi-permeable support in the bioreactors, while polycarbonate (PC) transwell® inserts were used in 12-well plates (Corning, New York, USA).

Prior to cell seeding, all the membranes were coated with a 1 μg/mL type I collagen solution from rat tails (Sigma Aldrich) and incubated overnight at 4°C. Then, cells were seeded directly in the bioreactor with a seeding density = $1 \times 10^5$ cells/cm$^2$ and cultured for 9 days in order to monitor the growth of the cellular layer. In the dynamic experiments, the bioreactors were connected to a peristaltic pump (ISMATEC, Cole-Parmer GmbH - Wertheim, Germany) 24 h after cell seeding to allow proper cell attachment to the membrane before being exposed to flow. The flow rate was maintained at 50 μL/min until the cells reached confluence at day 5, then it was set to 200 μL/min. Static controls were performed both in the TEEI bioreactor and in transwells® by seeding Caco-2 cells at the same density as used in dynamic experiments. Unless otherwise mentioned, all the reagents for cell culture were purchased from Sigma Aldrich (Missouri, USA).

TEEI (i.e. $|Z|$) measurements were acquired with the impedance-meter in the bioreactors both in a low and high frequency range (40 -1040 Hz and 2 kHz-100 kHz). While in the transwell, resistance measurements were performed with the EVOM (12.5Hz). To obtain TEER, the average of $|Z|$ values in the low frequency range and resistance values were subtracted from their respective blank ($|Z|$ or resistance in absence of cells) and normalised for the cell culture area (1.2 and 1.8 cm$^2$, respectively for transwells® and bioreactors).

Live imaging in the bioreactor was performed with an inverted microscope (Olympus - Tokyo, Japan) in bright field. At the end of the culture, cell viability was assessed with a resazurin based in metabolic assay (Sigma Aldrich). Then, cells were fixed and stained with rhodamine conjugated phalloidin, DAPI and occludin monoclonal antibody (OC-3F10) directly conjugated with Alexa Flour 488 (Thermo-Fisher, Massachusetts, USA) and images were acquired with a confocal microscope (Nikon A1, Tokyo, Japan).

**2.6 Statistical analyses**

All the experiments were performed in triplicate (n=3 independent experiments). The correlation between NaCl measurements performed with the cellular impedance-meter and with the EVOM was evaluated by calculating the Pearson correlation coefficient. Two-way ANOVA analysis was performed to compare TEER measurements under static and dynamic conditions at different time points (days of culture), performing multiple comparisons with the Tukey's test. One-way ANOVA analysis was used to evaluate the effect of flow on cell viability, using the Sidak test for multiple comparisons. Statistical analyses were performed in GraphPad Prism (GraphPad Software, San Diego, CA, USA), setting statistical significance at $p < 0.05$.

**3. Results**

**3.1 Computational models**

For a typical flow rate of 200 μL/min (Giusti et al., 2014), the shear stress on the membrane was between $8.0·10^{-6}$ and $4.3·10^{-7}$ Pa (Figure 3A). Moreover, the flow streamlines (in red) show that the flow on the cell culture membrane is laminar, without any turbulence or vortices.

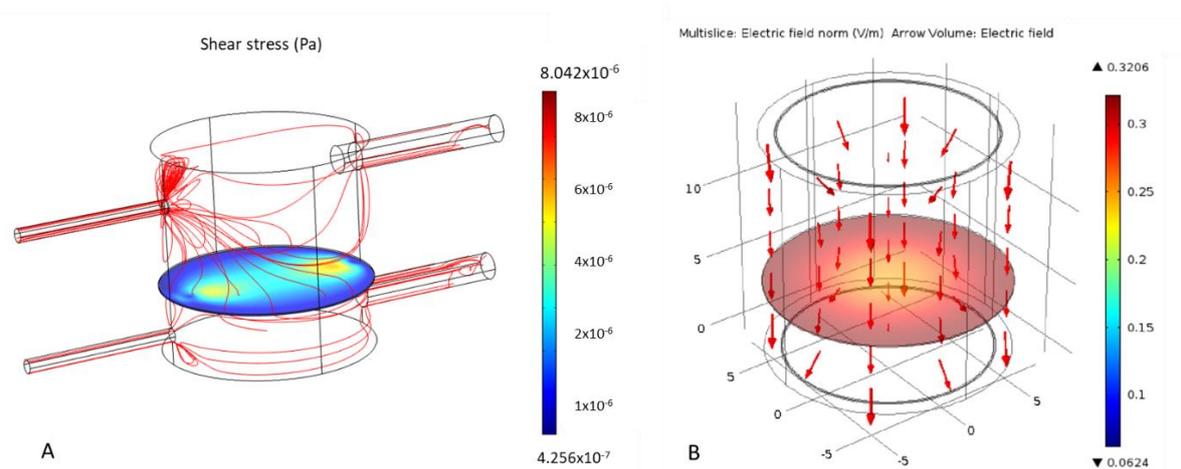

**Figure 3**: CFD model of the bioreactor A) Fluid-dynamics: red lines are the flow streamlines, while the colormap on the membrane represent the shear stress. B) FEM model of the electrical field distribution in the TEEI bioreactor: red arrows represent the direction of electric field, while the colourmap its intensity.

As highlighted by the arrows in **Error! Reference source not found.**3B, the electric field has a homogenous distribution within the TEEI bioreactor. More specifically, the electric field concentration is slightly higher (around 0.3 V/m) in the portion of the cellular layer underneath the electrodes, whereas at the center of the barrier the electric field values are around 0.25 V/m. The circular electrodes result in a much more homogeneous field than the EVOM chopsticks (schematized in Fig.1).

### 3.2 Preliminary tests with saline solutions

Preliminary tests were performed measuring the resistance and the impedance magnitude (|Z|) of NaCl solutions at different concentrations, respectively with EVOM in transwells® and with the impedance-meter in the bioreactors. Figure 4A shows the correlation between the low frequency impedance magnitude values in the TEEI bioreactor (|Z|, y-axis) versus the corresponding resistance values measured for the same solutions in transwells with the EVOM (|Z|$_{EVOM}$, x-axis). The Pearson coefficient was 0.99, showing good correlation between the two measurements.

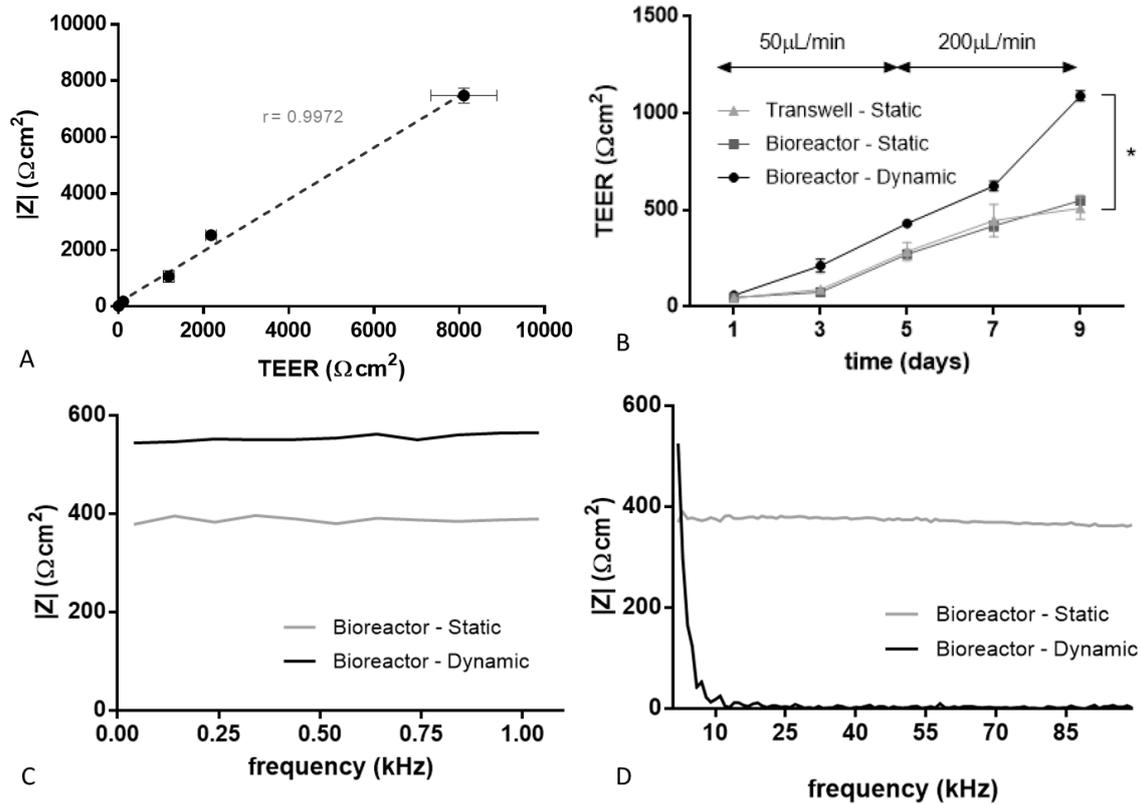

**Figure 4**: A) Comparison between NaCl measurements performed with the impedance-meter (y-axis) and the EVOM (x-axis). r = Pearson correlation coefficient; B) TEER measurements during cell culture performed with the impedance-meter (f<1 kHz) in the bioreactors and with the EVOM in the transwells (*=p<0.05 between static and dynamic conditions). TEEI measurement 5 days after seeding in the bioreactor in static and dynamic conditions: C) low frequencies (0.40- 1 kHz) and D) high frequencies (2-100 kHz).

### 3.4 Intestinal model in-vitro: real-time TEER/TEEI measurements and live imaging in the TEEI bioreactor

TEER measurements of a growing monolayer of Caco-2 cells cultured for 9 days in the TEEI bioreactor (both under static and dynamic conditions) and in Transwells (controls, static conditions) were performed with our impedance-meter and with the EVOM, respectively (Figure 4B). No significant differences were observed between TEER values measured over time under static conditions in the TEEI bioreactor and in Transwells, as expected. Conversely, the presence of flow significantly increases TEER values over time with respect to those obtained under static conditions. Moreover, the increase in TEER due to flow increases with culture time, as indicated by the 2-way ANOVA analyses which show significant interaction between time and flow. Examples of TEEI frequency sweeps are reported in Figures 4C-D, which show a comparison between the impedance modulus over frequency measured in the TEEI bioreactor after culturing cells for 5 days in static and dynamic conditions. In particular, in the dynamic case, at low frequencies (40-1040 Hz), the impedance magnitude |Z| remained constant, while, at high frequencies (2-100 kHz), it follows a typical RC circuit trend (i.e. current is passing through the paracellular pathway considering also the capacitive component of the cells (Benson et al., 2013)). On the contrary, in the static case, |Z| showed a constant trend for both the frequency ranges indicating the absence of a capacitive component.

In addition to TEER/TEEI measurements, the cell monolayer was monitored by optical imaging throughout the culture period (Figures 5A-C). Finally, to assess the organization of the cell cytoskeleton, fluorescence imaging was performed at the end of the experiment (Figures 5D-H). In particular, the presence of tight junctions was assessed with immunostaining for occludin, which is known to be a functional component of these multiprotein junctional complexes (McCarthy et al., 1996). Both the controls and the cells cultured under flow demonstrated high levels of peripheral occludin and actin expression, typical of differentiated Caco-2 cells. However, no relevant visual differences in occludin expression were observed between cells under static or dynamic conditions. On the contrary, it is possible to notice a neater aspect of cells cultured under flow (Figure 5D) respect to the static control (Figure 5E), where cell debris can be found on top of the cell monolayer. Notably, cell viability was significantly higher in dynamic conditions (**Error! Reference source not found.**5I) with respect to static conditions both in the transwells and in the bioreactors.

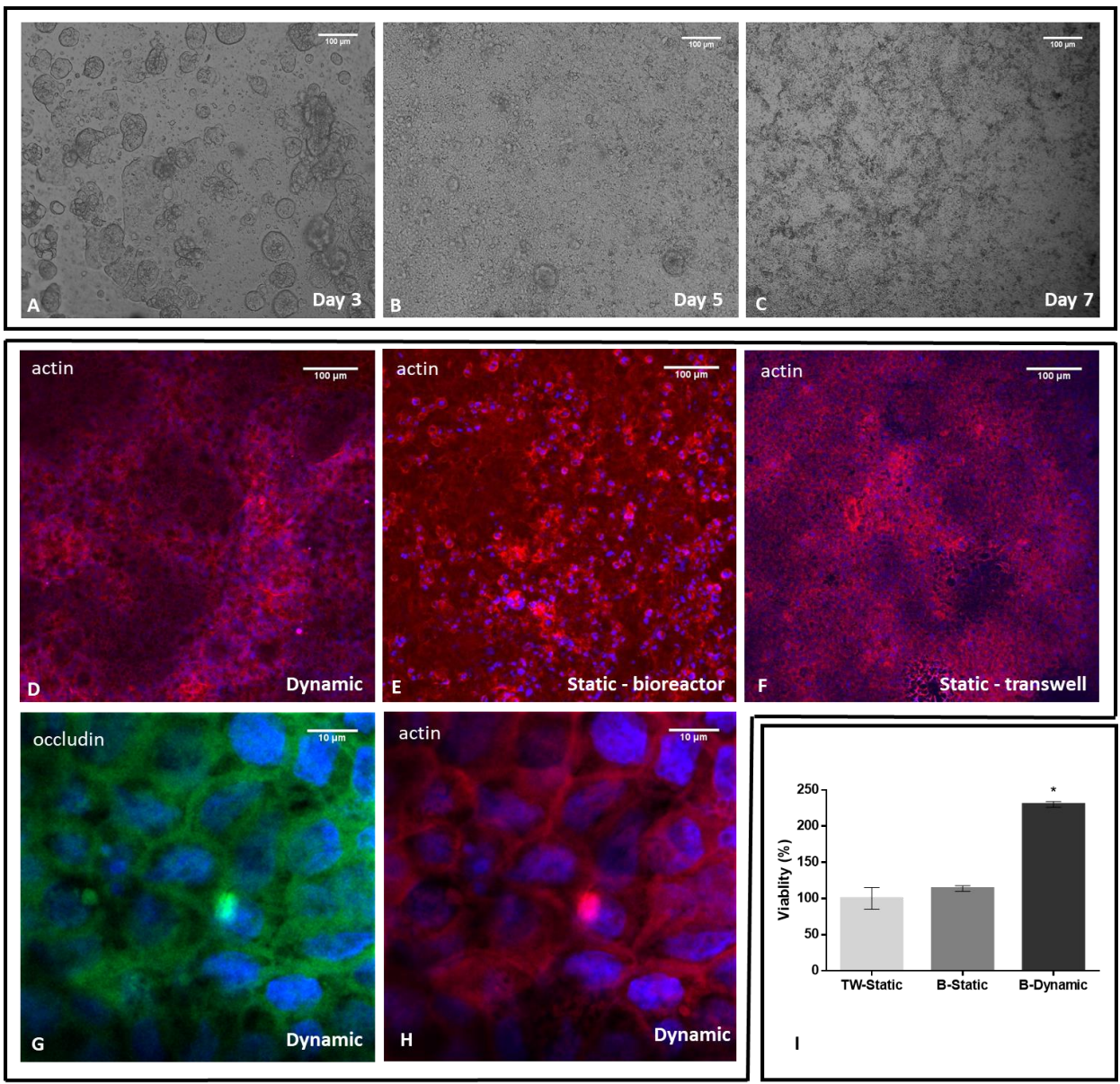

**Figure 5**: Live Imaging (10X bright field acquisition) inside the bioreactor under flow at day 3 (A), 5 (B) and 7 (C); Confocal Imaging at day 9 (nuclei are stained in blue and actin in red, 20X magnification): D) dynamic bioreactor, E) static bioreactor and F) transwell; G) Confocal Imaging at day 9 in dynamic conditions (40X magnification) - nuclei are stained in blue, occludin in green and H) actin in red; I) Viability at day 9 in the transwell and in the bioreactors (data are normalised with respect to static controls in transwells) *=p<0.05.

## 4. Discussion

In this paper we described the design and characterisation of a new dual flow TEEI bioreactor for continuous impedance monitoring and real-time imaging of biological barriers. Computed shear stress values in the bioreactor were lower than the typical physiological shear stresses reported in the literature for intestinal epithelia (~ $10^{-3}$ Pa) (Kim et al., 2012; Lentle R.G, Janssen, 2008). However, even lower values such as $6 \times 10^{-4}$ Pa (Giusti et al., 2014) or $1-5 \times 10^{-7}$ Pa (Costello et al., 2015) have been reported to have significant effects on the TEER values with respect to static controls. On the other hand, the FEM model of the electric field distribution in the device showed that it was homogenous at the membrane level (**Error! Reference source not found.**3B) with respect to the EVOM (**Error! Reference source not found.**), allowing a more precise electrical characterization of the cellular barrier. Furthermore, comparative experimental measurements of solution (NaCl) using the EVOM – currently considered the gold standard for TEER measurements – and our impedance measuring system (Figure 5A) confirmed its linearity and accuracy.

The system was therefore used to monitor the proliferation of Caco-2 cells in the TEEI bioreactor. The results reported in **Error! Reference source not found.**4B demonstrate that flow conditions improve TJ integrity as indicated by the consistently higher TEER values throughout the culture period. In addition, optical observation through the bioreactor imaging windows showed that cell confluency was more rapid under flow, indicating that dynamic conditions stimulate cell proliferation. In general, TEER values of around 400 $\Omega cm^2$ are indicative of a confluent Caco-2 monolayer (Srinivasan et al., 2015); this value was reached in 5 days in the TEEI bioreactor under flow while under static conditions it was reached at day 7 (in both the Transwell and static bioreactor controls). Moreover, the increase in TEER after confluence was limited under static conditions and seemed to reach a plateau after day 7. On the contrary, under dynamic conditions TEER values showed an increasing trend until day 9. This result is in agreement with the fact that shear stress is associated with a mechano-transductive effect, in particular the molecular pathways that stimulate the formation of TJs are activated as a consequence of cell exposure to flow (Giusti et al., 2014; Kim et al., 2012; Srinivasan et al., 2015).

One of the main advantages offered by the TEEI bioreactor is the possibility to analyse the whole impedance frequency spectrum from 40 Hz up to 100 kHz (Figure 4C-D) throughout the cell culture period. This allows more accurate characterisation of barrier formation with respect to the evaluation of TEER values, as the latter can be influenced by additional factors (e.g. cell passage number, etc) as well as TJ formation (Man et al., 2008). Moreover, thanks to both bright field and fluorescence imaging (Figure 5A-C), the formation of dome structures, typical of Caco-2 and other epithelial cells (Gorrochategui et al., 2016; Grasset et al., 1984; Hara et al., 1993; MacCallum et al., 2005; Mariadason et al., 2000) can be observed. These are more evident under dynamic conditions. Their formation is associated with cell polarization and the presence of 'functional tight junctional complexes' and are indicative of epithelial differentiation (Mariadason et al., 2000) (Terrés et al., 2003). Finally, the cell metabolic assay (Figure 5H), show that dynamic conditions not only improve the tightness of the barrier but also cellular viability.

## 5. Conclusions

The TEEI bioreactor combines integrated sensing with fluidics for the generation of advanced in-vitro models of biological barriers. Besides demonstrating the feasibility of the system, this study demonstrates that providing shear stress is fundamental for developing physiologically relevant in-vitro barrier models. The proposed bioreactor can thus be powerful tool for dynamic cell cultures, allowing the monitoring of the cells through real-time impedance measurements and live-imaging.

## Acknowledgments

The study has received funding from the European Union's Horizon 2020 research and innovation programme under grant agreement No 760813 (PATROLS).

endothelial barriers in human disease. Drug Discov. Today 10, 395–408. https://doi.org/10.1016/S1359-6446(05)03379-9

Onnela, N., Savolainen, V., Juuti-Uusitalo, K., Vaajasaari, H., Skottman, H., Hyttinen, J., 2012. Electric impedance of human embryonic stem cell-derived retinal pigment epithelium. Med. Biol. Eng. Comput. 50, 107–116. https://doi.org/10.1007/s11517-011-0850-z

Peng, Y., Yadava, P., Heikkinen, A.T., Parrott, N., Railkar, A., 2014. Applications of a 7-day Caco-2 cell model in drug discovery and development. Eur. J. Pharm. Sci. 56, 120–130. https://doi.org/10.1016/j.ejps.2014.02.008

Sbrana, T., Ucciferri, N., Favr??, M., Ahmed, S., Collnot, E.M., Lehr, C.M., Ahluwalia, A., Liley, M., 2016. Dual flow bioreactor with ultrathin microporous TEER sensing membrane for evaluation of nanoparticle toxicity. Sensors Actuators, B Chem. 223, 440–446. https://doi.org/10.1016/j.snb.2015.09.078

Srinivasan, B., Kolli, A.R., Esch, M.B., Abaci, H.E., Shuler, M.L., Hickman, J.J., 2015. TEER Measurement Techniques for In Vitro Barrier Model Systems. J. Lab. Autom. 20, 107–126. https://doi.org/10.1177/2211068214561025

Terrés, A.M., Windle, H.J., Ardini, E., Kelleher, D.P., 2003. Soluble extracts from Helicobacter pylori induce dome formation in polarized intestinal epithelial monolayers in a laminin-dependent manner. Infect. Immun. 71, 4067–4078. https://doi.org/10.1128/IAI.71.7.4067-4078.2003

Unzel, D.G.¨, Zakrzewski, S.S., Schmid, T., Pangalos, M., Wiedenhoeft, J., Blasse, C., Ozboda, C., Krug, S.M., 2012. From TER to trans-and paracellular resistance: lessons from impedance spectroscopy. Ann. N.Y. Acad. Sci 1257, 142–1512. https://doi.org/10.1111/j.1749-6632.2012.06540.x

Vogel, P. a., Halpin, S.T., Martin, R.S., Spence, D.M., 2011. Microfluidic transendothelial electrical resistance measurement device that enables blood flow and postgrowth experiments. Anal. Chem. 83, 4296–4301. https://doi.org/10.1021/ac2004746

Wegener, J., Keese, C.R., Giaever, I., 2000. Electric cell-substrate impedance sensing (ECIS) as a noninvasive means to monitor the kinetics of cell spreading to artificial surfaces. Exp. Cell Res. 259, 158–166. https://doi.org/10.1006/excr.2000.491917